\documentclass[%
 reprint,
 amsmath,amssymb,
prl,
]{revtex4-1}

\usepackage{graphicx}
\usepackage{dcolumn}
\usepackage{braket}
\usepackage{bm}
\usepackage{tikz}

\newcommand{\tikzcircle}[2][orange,fill=orange]{\tikz[baseline=-0.5ex]\draw[#1,radius=#2] (0,0) circle ;}%
\begin{document}


\title{Rydberg interaction induced enhanced excitation in thermal atomic vapor}

\author{Dushmanta Kara$^{1}$}
\email{dushmantakara1@niser.ac.in}%
\author{Arup Bhowmick$^{1}$}
\author{Ashok K. Mohapatra$^{1}$}%
 \email{a.mohapatra@niser.ac.in}
\affiliation{%
$^{1}$School of Physical Sciences, National Institute of Science Education and Research Bhubaneswar, HBNI, Jatni - 752050, India}

\begin{abstract}
{\bf We present the experimental demonstration of interaction induced enhancement in Rydberg excitation or Rydberg anti-blockade in thermal atomic vapor. We have used optical heterodyne detection 
technique to measure Rydberg population due to two-photon excitation to the Rydberg state. The anti-blockade peak which doesn't satisfy the two-photon 
resonant condition is observed along with the usual two-photon resonant peak which can't be explained using the model with non-interacting three-level atomic system.  A model involving two 
interacting atoms is formulated for thermal atomic vapor using the dressed states of three-level atomic system to explain the experimental observations. A non-linear dependence of vapor density 
is observed for the anti-blockade peak which also increases with increase in principal quantum number of the Rydberg state. A good agreement is found between the experimental observations and the proposed 
interacting model. Our result implies possible applications towards quantum logic gates using Rydberg anti-blockade in thermal atomic vapor.}

\end{abstract}


\maketitle
 
Long range many body interaction in Rydberg atoms give rise to many interesting phenomena. The suppression in Rydberg population or the excitation blockade is the most striking one giving rise to a 
variety of applications~\cite{saff10}. A highly dense atomic ensemble behaves like a single super atom producing a strongly correlated many body system~\cite{dudi_112} and also leading to a single photon source~\cite{dudi_212}. The phenomenon has been experimentally observed in an atomic ensemble in a magneto optical trap~\cite{tong04, sing04, lieb05, vogt06}, in a magnetic trap~\cite{heid07, rait08} and also in a single atom trap~\cite{urba09,gaet09}. Many theoretical models focus on study of strongly correlated many body system in ultra cold atom and Bose Einstein condensate~\cite{weim08, pohl10, pupi10, henk10}. Rydberg blockade interaction can induce optical non-linearity which is non-local~\cite{busc17} and also strong enough to observe for single photon~\cite{peyr12,fist13}. Rydberg blockade interaction may also leads to applications such as quantum gates using atoms~\cite{jaks00, isen10, wilk10, luki01}. An opposite effect of Rydberg blockade with enhancement in Rydberg excitation facilitated by  interaction or Rydberg anti-blockade has been proposed in ultra cold atomic gas  using a two photon excitation to Rydberg state~\cite{ates07}. An experiment  performed in ultra cold ensemble of atom verifies the effect based on the theoretical model~\cite{amth10}. It has been proposed that resonant dipole dipole interaction has non-additive character due to anti-blockade in an ensemble having more than two atoms in a the blockade sphere~\cite{pohl09}. In addition to this, the existence of anti-blockade between two Rydberg atoms, interacting with a zero area phase jump pulse is also reported~\cite{qian09}. The implementation of quantum logic gate using Rydberg anti-blockade has also recently been proposed~\cite{sark15,su16,su17}.  

Recent experiments with thermal vapor have drawn the attentions to study Rydberg  interaction induced manybody effect~\cite{balu13,klei17, melo16}. Electromagnetically induced transparency involving Rydberg state in thermal vapor cell as well as in micron size vapor cell has been studied~\cite{moha07,kubl10}. In addition four wave mixing for a Rydberg state~\cite{koll12} and kerr non-linearity in Rydberg EIT has also been reported in thermal Rubidium vapor~\cite{bhow16}. A recent study of Rydberg blockade in thermal atomic vapor has also been performed~\cite{bhow_217}.  Anomalous excitation facilitated by Rydberg interaction has also been proposed recently in thermal atomic vapor~\cite{lets17}. 

In this article, we present a strong evidence of enhancement on Rydberg excitation due to interaction in thermal atomic vapor. A two atom interacting model is formulated using the dressed state picture of a three level system in cold atomic ensemble. The model is further extended to thermal atomic vapor by Doppler averaging it over the ensemble. An experiment have been performed in thermal rubidium vapor using optical heterodyne detection technique~\cite{bhow_117} to observe the effect. A good match is found between the model and the experiment observation as an evidence of the existence of the anti-blockade in thermal atomic vapor.

\textbf{Theoretical Model.} Let us consider a three level atomic system with states $\ket{g}$, $\ket{e}$ and $\ket{r}$ as shown in Fig. 1(a). The probe (coupling) laser drives the transition 
$\ket{g}\rightarrow \ket{e}$ ($\ket{e}\rightarrow \ket{r}$) with rabi frequency $\Omega_{P}$ ($\Omega_{C}$) and detuning $\Delta_{P}$ ($\Delta_{C}$). Both of these transitions are dipole allowed 
whereas the transition $\ket{g}\rightarrow \ket{r}$ is dipole forbidden. The population decay rate of the transitions $\ket{e}\rightarrow\ket{g}$ and $\ket{r}\rightarrow\ket{e}$ are $\Gamma_{eg}$ and 
$\Gamma_{re}$ respectively. We have also included a decay rate $\Gamma_{rg}$ to account the indirect decay of the state $\ket{r}\rightarrow\ket{g}$ due to finite transit time of the 
thermal atoms. In the regime $\Omega_{C}\ll\Omega_{P}$, interaction with the coupling laser can be treated as a small perturbation to the atomic states dressed by the strong probe laser.
For large probe detuning $\Delta_{P}\gg\Omega_{P}$ and $\Gamma_{eg}$, the dressed states are given by $\ket{g_{1}}\approx\ket{e}+\frac{\Omega_{P}}{2\Delta_{P}}\ket{g}$ and 
$\ket{g_{2}}\approx\ket{g}-\frac{\Omega_{P}}{2\Delta_{P}}\ket{e}$ with difference in their energy eigenvalues $\Delta\simeq\Delta_{P}+\Omega_{P}^{2}/2\Delta_{P}$.
The steady state population of the dressed states can be determined by diagonalizing the steady state density matrix for the two-level atomic transition $\ket{g}\rightarrow\ket{e}$ driven by the strong
probe laser. The population of the states $\ket{g_{1}}$ and $\ket{g_{2}}$ are found to be approximately $\Omega_{P}^{4}/16\Delta_{P}^{4}$ and $1$ respectively. When the coupling laser is scanned over these dressed states, each of them will behave like a ground state exciting to the Rydberg state by the coupling laser. It is to be noted that the optical pumping rate to achieve the steady state population 
of the dressed states is $\Gamma_{eg}$. If $\Omega_{C}\ll\Gamma_{eg}$ then the coupling laser driving to the Rydberg state can not build the coherence between the dressed states. Hence, both the 
states can be treated independently and the total Rydberg population can be determined by adding the individual Rydberg populations driven from each of the dressed states. The coupling Rabi frequencies are 
scaled for each dressed state as $\Omega_{1}\approx\Omega_{C}$ for the transition $\ket{g_{1}}\rightarrow\ket{r}$ and $\Omega_{2}\approx\Omega_{P}\Omega_{C}/2\Delta_{P}$ for the transition
$\ket{g_{2}}\rightarrow\ket{r}$. Similarly, the population of the Rydberg state will decay to the states $\ket{g_{1}}$ and $\ket{g_{2}}$ with decay rates as
$\Gamma_{1}\simeq \Gamma_{re}+\dfrac{\Omega_{P}}{2\Delta_{P}}\Gamma_{rg}$ and $\Gamma_{2}\simeq\Gamma_{rg}+\dfrac{\Omega_{P}}{2\Delta_{P}}\Gamma_{re}$ respectively. 
   
\begin{figure}[t]
\includegraphics[scale=0.4]{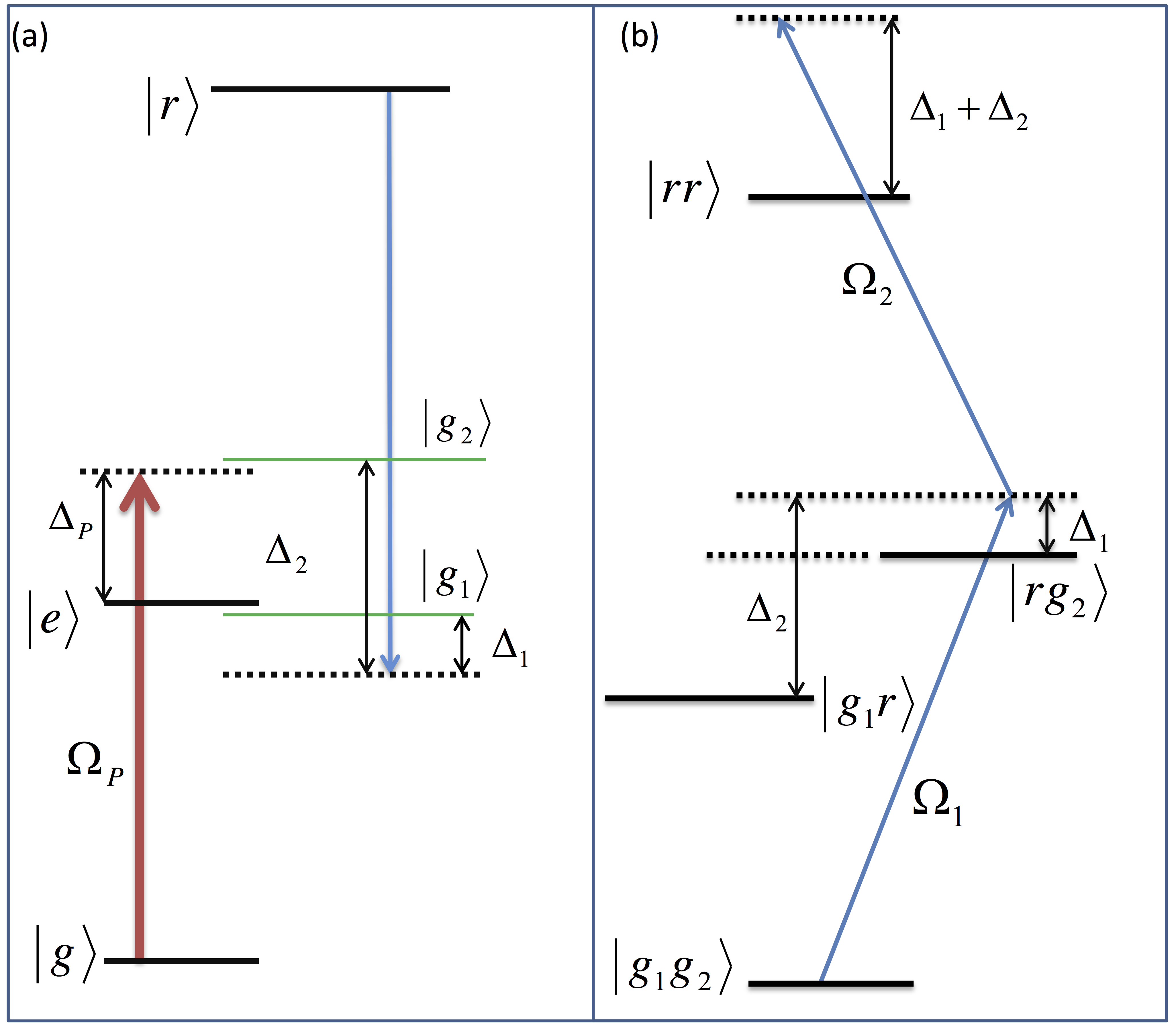}
\caption{\label{fig:model1} The relevant energy level diagrams. (a) A probe laser driving the $\ket{g}\rightarrow \ket{e}$ transition of a single atom leads to the dressed states $\ket{g_1}$ and 
$\ket{g_2}$. The coupling laser drives the dressed states to the Rydberg state $\ket{r}$ with detuning $\Delta_{1}\simeq\Delta_{C}-\Omega_{P}^{2}/4\Delta_{P}$ and $\Delta_{2}\simeq\Delta_{P}+\Delta_{C}+\Omega_{P}^{2}/4\Delta_{P}$ respectively. (b) The two atom model with states $\ket{g_{1}g_{2}}$ representing an atom in each
dressed states, $\ket{g_{1}r}$, $\ket{rg_{2}}$ representing one atom in respective dressed state and other atom in Rydberg state and $\ket{rr}$ representing both the atoms in Rydberg state.}
\end{figure}

Considering two atoms driven to the Rydberg state simultaneously, there are four possible dressed states as $\ket{g_{1}g_{1}}$, $\ket{g_{2}g_{2}}$, $\ket{g_{1}g_{2}}$ and $\ket{g_{2}g_{1}}$. 
The states $\ket{g_{1}g_{2}}$ and $\ket{g_{2}g_{1}}$ are degenerate and have equal steady state 
population of $\Omega_P^4/16\Delta_P^4$. The population of the state $\ket{g_2g_2}$ is approximately $1$ and of $\ket{g_{1}g_{1}}$ is negligibly small for $\Omega_P\ll\Delta_P$. 
The difference in the energy eigenvalues of $\ket{g_2g_2}$ and $\ket{g_1g_2}$ is $\Delta$ which can be made larger than the typical Doppler linewidth of the transition in thermal vapor. 
If a narrow band laser is made resonant to the transition $\ket{g_1g_{2}}\rightarrow\ket{rg_2}$ then the laser will be out of resonance to drive the atoms in the state $\ket{g_{2}g_{2}}$ to the 
Rydberg state. In the regime $\Omega_C\ll\Gamma_{eg}$, the coupling laser can't introduce coherence between the states $\ket{g_1g_2}$ and $\ket{g_2g_1}$. Hence, either of the states can be 
considered in the two atom model to determine the Rydberg population with proper normalisation accounting for both the states. Thus in the simplified model, only one of the dressed states of the two atomic system  can be considered to model the anti-blockade peak. As shown in figure 1(b), the relevant energy level diagram to model the anti-blockade peak are $\ket{g_{1}g_{2}}$, $\ket{g_{1}r}$, $\ket{rg_{2}}$ and $\ket{rr}$.    

The Hamiltonian for the two atoms is given by $H=H^{(1)}\otimes\mathbf{I}+\mathbf{I}\otimes H^{(2)}$ where $H^{(1)}$ and $H^{(2)}$ are the Hamiltonian of atom 1 and atom 2, given by $H^{(1)} =\dfrac{-\hbar}{2}( \Delta \ket{g_{1}}\bra{g_1} + \Delta_{1}\ket{r}\bra{r} + \Omega_1\ket{g_1}\bra{r} + H.C.)$ and
$H^{(2)}=\dfrac{-\hbar}{2}\left(\Delta_{2}\ket{r}\bra{r}+\Omega_2\ket{g_2}\bra{r}+H.C.\right)$ respectively. The Lindblad operator for the two-atom model can be written as
$\mathcal{L}_{D}=\mathcal{L}_{D_{1}}\otimes \rho^{(2)}+\rho^{(1)}\otimes\mathcal{L}_{D_{2}}$, where 
$\mathcal{L}_{D_{j}=}\Gamma_j\left(C\rho^{(j)}C^{\dagger}-\dfrac{1}{2}(C^{\dagger} C\rho^{(j)}+\rho^{(j)}C^{\dagger} C)\right)-\dfrac{\Gamma_{eg}}{2}(C^{\dagger}C\rho^{(j)}C C^{\dagger}+ H.C.)$, where $j=1,2$ for atom 1 and atom 2 respectively and $C=\ket{f}\bra{i}$, where $\ket{i}$ stands for the initial state from which the population decays to the the final state $\ket{f}$. $\rho^{(1)}$ and $\rho^{(2)}$ represents the density matrices of atoms 1 and 2 respectively~\cite{begu13}.
Rydberg population in the system can be determined from the steady state solution of the master equation $\dot{\rho}=\dfrac{1}{i\hbar}[H,\rho]+\mathcal{L}_{D}$ where $\rho=\rho^{(1)}\otimes\rho^{(2)}$ for 
non-interacting atoms. Rydberg population can be determined as $\rho_{rr}=(\dfrac{\rho_{22}+\rho_{33}}{2}+\rho_{44})\dfrac{\Omega_{P}^{4}}{8\Delta_{P}^{4}}$ where 
the factor $\dfrac{\Omega_{P}^{4}}{8\Delta_{P}^{4}}$ is the population of the state $\ket{g_{1}g_{2}}$ including a normalization factor of 2 to account for the population of the state $\ket{g_2g_1}$

\begin{figure}[t]
\includegraphics[scale=0.3]{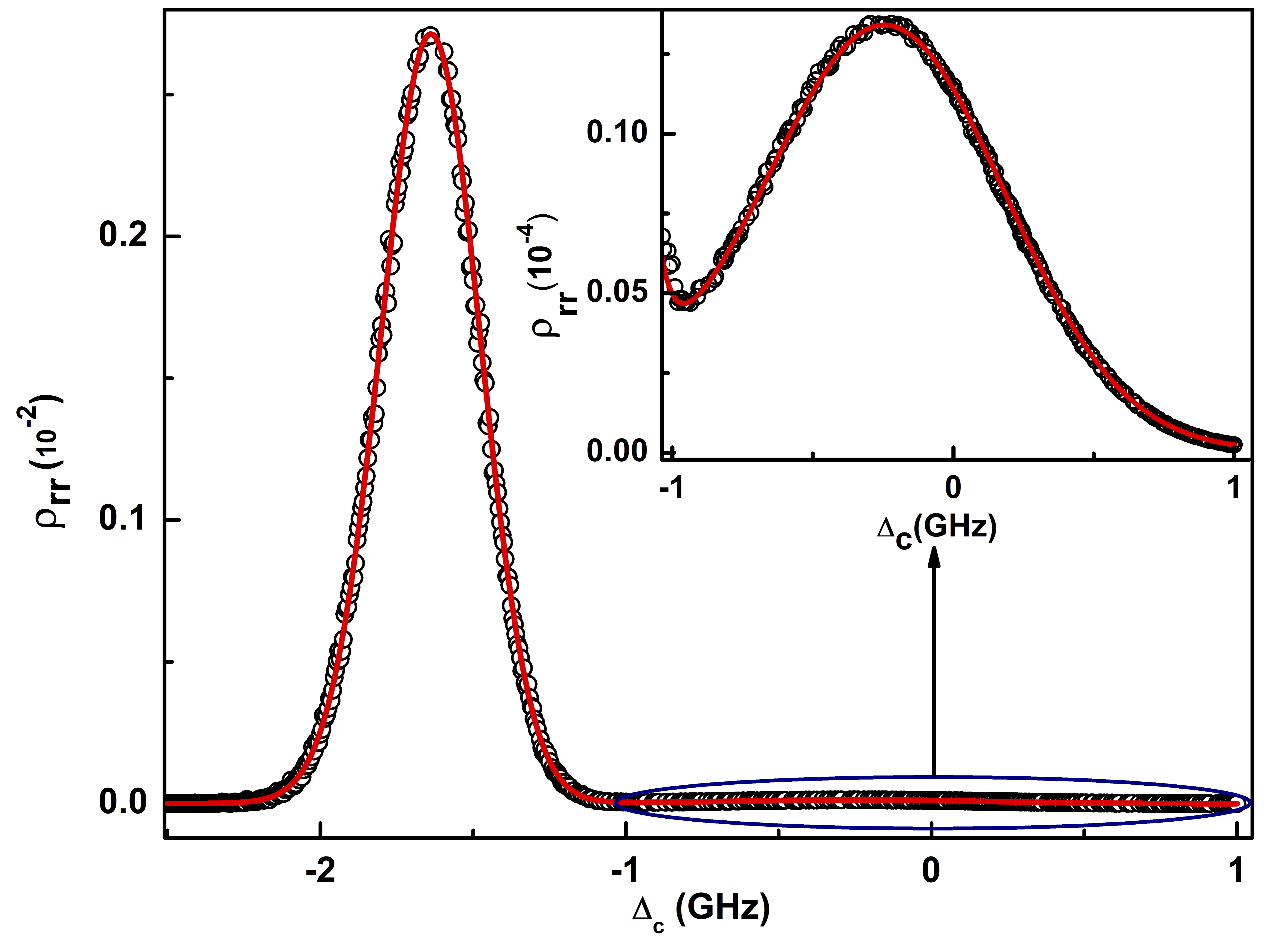}
\caption{\label{fig2} Rydberg population as a function of coupling laser detuning. Rydberg population calculated using exact three level single atomic system and two-atoms non-interacting model are 
represented by the solid line and the symbol ($\circ$) respectively. Inset shows the magnified view of the peak near $\Delta_C=0$. Laser parameters used in the calculation are $\Omega_{P}=400$ MHz, $\Omega_{C}=5$ MHz and $\Delta_{P}=1.25$ GHz.}
\end{figure}

To model the system for thermal atomic vapor, consider both the atoms moving with independent velocity $v_{l}$ where $l$-index takes the value $1$ or $2$ depending on the atom being considered. For counter-propagating 
configuration of the probe and coupling lasers with wave vector $k_P$ and $k_C$ respectively, the detunings are modified as $\Delta_{P}=\Delta_{P}-k_{P}v_{l}$ and 
$\Delta_{C}=\Delta_{C}+k_{C}v_{l}$ with $\Delta k=k_{C}-k_{P}$. Thus the laser detunings from the two-atom transitions are modified as 
$\Delta_{1}=\Delta_{C}+k_{C}v_{1}-\Omega_{P}^{2}/4(\Delta_{P}-k_{p}v_{1})$ and $\Delta_{2}=\Delta_{P}+\Delta_{C}+(\Delta k)v_{2}+\Omega_{P}^{2}/4(\Delta_{P}-k_{p}v_{2})$.  Doppler averaged Rydberg population can be determined by calculating the integral $\rho_{rr}=\dfrac{1}{\pi v_{p}^{2}}\int\int \rho_{rr}(v_{1},v_{2})e^{-v_{1}^{2}/v_{p}^{2}}e^{-v_{2}^{2}/v_{p}^{2}}dv_{1}dv_{2}$ with $v_{p}$ being the most probable velocity. 
We have used a Monte-Carlo technique to evaluate the integral. The comparison of the non-interacting two-atom model with the exact single atom calculation is depicted in figure~\ref{fig2}. An excellent
match is found for $\Omega_C\ll\Gamma_{eg}$, which tends to deviate with increase in $\Omega_C$ above $\Gamma_{eg}$. The main peak at two-photon resonance has to be calculated 
using the two-atom model by driving the atoms from the state $\ket{g_2g_2}$ to the Rydberg state.

\begin{figure}[t]
\includegraphics[scale=0.4]{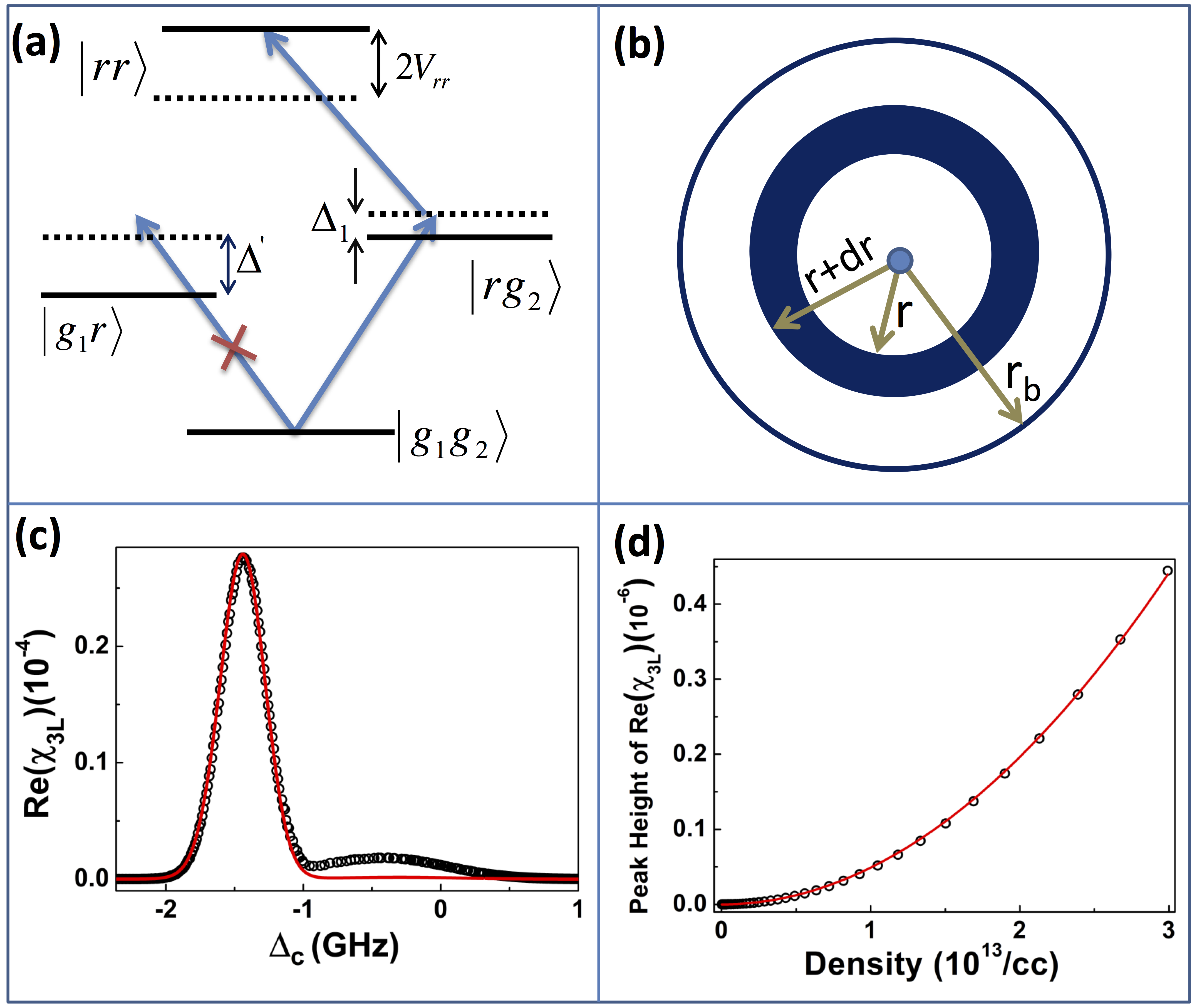}
\caption{\label{fig3}  (a) Energy level diagram to model Rydberg anti-blockade.  $\Delta_1$ and $\Delta_2$ $(=\Delta_1+\Delta^{\prime})$ are the coupling laser detuning of atom 1 and 2 respectively. (b) Schematic of the interaction sphere of radius $r_b$. The atom in the dressed state $\ket{g_2}$ is placed at the centre
of the sphere. The atoms in the spherical shell of radius $r$ and thickness $dr$ can compensate for the two-photon resonance to the $\ket{rr}$ state due to van der Waals type Rydberg-Rydberg interaction ($V_{rr}=C_6/r^6$)
with the atom at the center. (c) Dispersion spectrum of the probe laser calculated using non-interacting model (solid line) and two-atoms interacting model (\tikzcircle[black]{2pt}). 
(d) Calculated dispersion peak height of the anti-blockade peak using two-atoms interacting model (\tikzcircle[black]{2pt}) showing the quadratic dependance (solid line) of density.}
\end{figure}

Rydberg-Rydberg interaction can easily be introduced in the model by including the shift in energy of the $\ket{rr}$ state. Referring to figure~\ref{fig3}(a), consider a case where the narrow band laser is  
resonant to the $\ket{g_{1}g_{2}}\rightarrow\ket{rg_{2}}$ transition. Then the atom in the dressed state $\ket{g_1}$ is excited to the Rydberg state. The same laser will be out of resonance to the
$\ket{g_{1}g_{2}}\rightarrow\ket{g_{1}r}$ transition.  If the Rydberg-Rydberg interaction is absent then the state $\ket{rr}$ will also not satisfy the resonant condition. Therefore  the 
second atom in the dressed state $\ket{g_2}$ can't be excited to Rydberg state. Suppose the interaction shift of the $\ket{rr}$ state is equal to $\Delta^{\prime}$ 
(difference in the resonant frequencies corresponding to the $\ket{g_1g_2}\rightarrow\ket{rg_2}$ and $\ket{g_1g_2}\rightarrow\ket{g_1r}$ transitions) then the $\ket{rr}$ 
state will be resonant to the laser as shown in figure 3(a). Now the second atom present in the state $\ket{g_{2}}$ will also be excited to the Rydberg state unlike a non-interacting system. So the presence of the Rydberg interaction facilitate the excitation of the second atom enhancing the total Rydberg population compared to the non-interacting case and this phenomenon is known as 
Rydberg anti-blockade. Rydberg anti-blockade peak appears when the coupling laser is resonant to the $\ket{g_1}\rightarrow\ket{r}$ transition i.e. near $\Delta_C=0$ whereas the usual two-photon resonant
peak appears with the coupling laser resonating to the $\ket{g_2}\rightarrow\ket{r}$ transition. 

Rydberg population due to anti-blockade in thermal vapour was calculated as explained below. Referring to figure 3(b), consider an interaction sphere with radius $r_{b}$ where $r_b$ is defined as the blockade radius, which is given by $r_b=\sqrt[6]{\frac{C_6}{h\Omega_2}}$, with $C_6$ being the coefficient of van der Waals interaction. Consider the atom in the dressed state $\ket{g_1}$ 
resonating to the coupling laser to be at the center of the sphere. Assume that the second atom in the dressed state $\ket{g_2}$ is present in a concentric spherical shell with radius $r$ and 
thickness $dr$. Referring to figure 3(a), if $\Delta_1+\Delta^{\prime}=2C_6/r^6$, then the atom will be excited to the $\ket{rr}$ state which will lead to the enhanced
Rydberg excitation or anti-blockade. In a thermal atomic ensemble, effect of the atomic velocity distribution should be included in the detuning $\Delta_1$ and also in $\Delta^{\prime}$. 
Hence, in thermal atomic vapor, not all but only a specific velocity class of atoms in the given spherical shell of radius $r$ can satisfy the resonance condition to the $\ket{rr}$ state.
The resonance condition to the $\ket{rr}$ state constrain the velocity of the second atom to depend on the velocity of first atom as well as their inter-particle separation $r$. If the resonant atoms within the line width of Rabi coupling are assumed to contribute significantly to the anti-blockade then the above constraint can be used to reduce the complexity of the model. Taking the vapor density as $\eta$,  the number of atoms in the dressed state $\ket{g_2}$ present inside the spherical shell with radius $r$ is $\eta4\pi r^2 dr$. Suppose, only the velocity class of atoms at $v_2$ within a small velocity 
width $\Delta v_2=\Omega_2/\Delta k$ inside the same spherical shell satisfy the resonance condition to the $\ket{rr}$ state, then the effective number of atoms inside the interaction sphere 
contributing to the anti-blockade can be evaluated as  $N_{b}(v_1)=\frac{4\pi \eta}{\sqrt{\pi}v_{p}}\frac{\Omega_2}{\Delta k}\int_0^{r_{b}} r^2e^{-v_2^2/v_p^2} dr$.    
In the case of probe laser detuning larger than the Doppler width, $\Delta_P\gg k_Pv_2$, the light shift of state $\ket{g_2}$ can be expanded to be $\Omega_P^2/4\Delta_P(1+k_Pv_2/\Delta_P)$ neglecting the higher order terms. Using this the velocity of the second atom is found to be $v_{2}=(2C_6/r^6-\Delta_{1}^{'}({v_{1}}))/\Delta k^{'}$, where $\Delta_{1}^{'}({v_{1}})=\Delta_{P}+2\Delta_{C}+k_{c}v_{1}-{\Omega_{P}^{2}k_{P}v_{1}}/{4\Delta_{P}^{2}}$ and $\Delta k^{'}=\Delta{k}+\Omega_{P}^{2}k_{P}/4\Delta_{P}^{2}$ . 
The above integral can be solved analytically to find the total no of atoms contributing to anti-blockade as
$N_{b}(v_{1})=\frac{\pi \eta \Omega_{2}\sqrt{8C_{6}}\Delta k^{'}}{\Delta k (\Delta_{1}^{'}({v_{1}}))^{\frac{3}{2}}}$. Rydberg population can be related to the dispersion which is a measurable quantity using optical heterodyne detection technique as 
$\Re\left(\chi_{3L}(v_1)\right)=\dfrac{\eta\mid\mu_{eg}\mid^{2}N_{b}(v_{1})}{\epsilon_{0}\hbar(\Delta_{P}-k_pv_1)}\rho_{rr}(v_1)$~\cite{bhow16}. Averaging over the velocity distribution of the first atom, the dispersion
of the probe due to anti-blockade can be evaluated as
\begin{widetext} 
\begin{equation}
\Re\left({\chi_{3L}}\right)=\dfrac{\Omega_{2}\sqrt{8\pi C_{6}}\Delta k^{'}\mid\mu_{eg}\mid^{2}}{\epsilon_{0}\hbar v_{p}\Delta k}\eta^2 \int^{\infty}_{-\infty}\dfrac{\rho_{rr}(v_{1})e^{({-v_{1}^{2}}/{v_{p}^{2}})}}{(\Delta_P -k_P v_{1})(\Delta_{1}^{'}({v_{1}}))^{\frac{3}{2}}}dv_{1}
\end{equation}  
\end{widetext}
The probe dispersion spectrum by changing the coupling laser detuning was calculated from the theoretical model which is shown in fig. 3(c). However the model will explain only the anti-blockade peak as the laser was resonant to the transition $\ket{g_1 g_2}\rightarrow\ket{rg_2}$. The usual two photon resonance peak can be determined by considering excitation from the ground state $\ket{g_2g_2}$. For a comparison, the exact 3-level single atom calculation is also depicted in the same figure. Anti-blockade peak is observed to be enhanced significantly due to interaction compared to the non-interacting case. Referring to equation (1) the dispersion of the probe beam due to two atom interaction depends strongly on the principal quantum number of the Rydberg state and also depends quardratically on the density of the atomic vapor. The dispersion peak height of the anti-blockade peak calculated from the interacting model showing a quadratic dependence on the density of the atomic vapor is depicted in fig. 3(d).

\begin{figure}[t]
\includegraphics[scale=0.65]{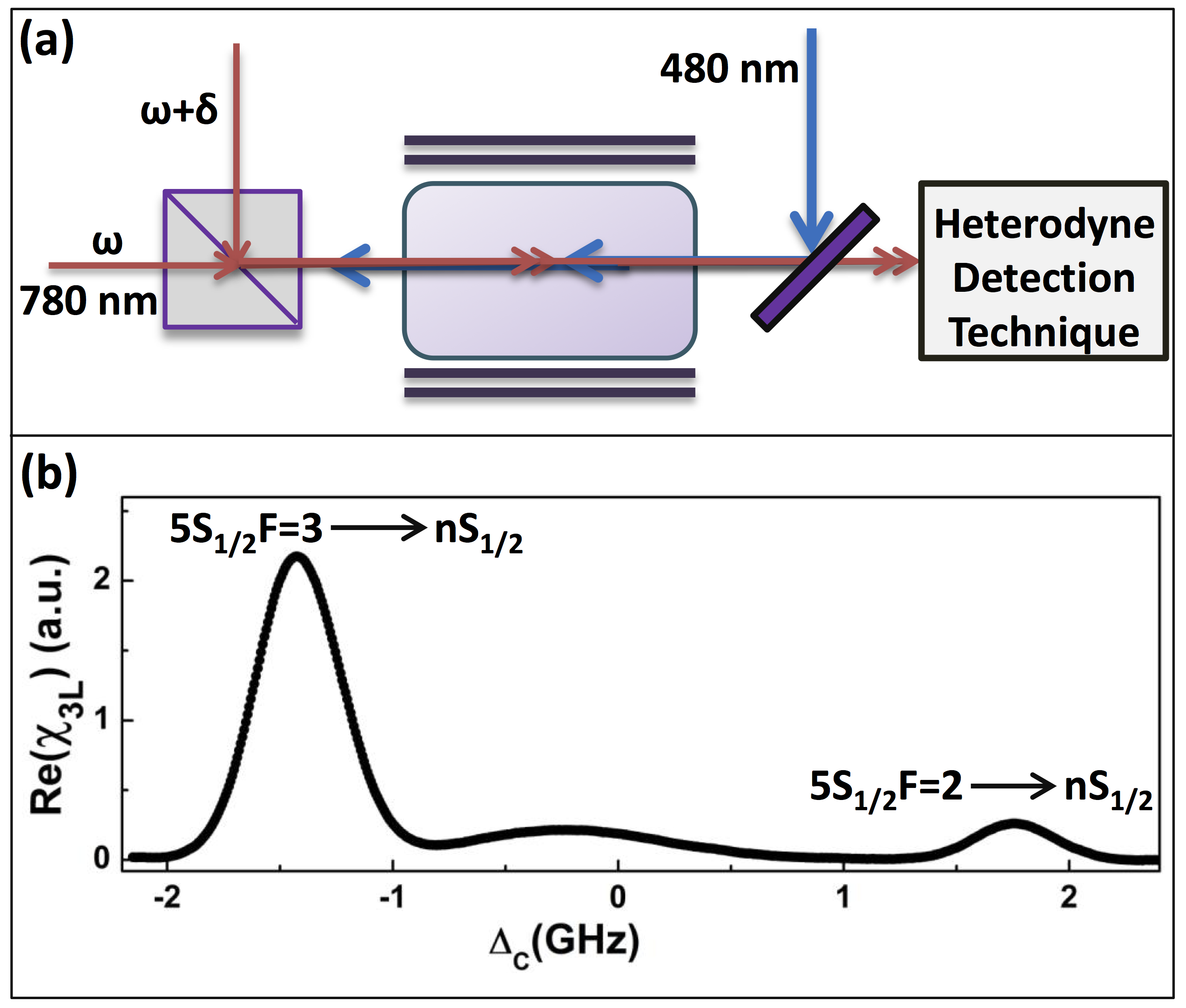}
\caption{\label{fig4} (a) Schematic of the experimental set up. (b) A typical dispersion spectrum of the probe laser observed using the optical heterodyne 
detection technique showing the resonance peaks corresponding to $5$s$_{1/2}$ F$=3\rightarrow n$ s$_{1/2}$ and $5$s$_{1/2}$ F$=2\rightarrow n$ s$_{1/2}$ of $^{85}$Rb.}
\end{figure} 

\textbf{Experimental results and discussions.} Schematic of the experimental set up is depicted in figure 4(a). Optical heterodyne detection technique (OHDT)~\cite{bhow16,bhow_117} was used to measure the dispersion of the probe beam 
propagating through a magnetically shielded rubidium vapor cell. The technique requires a probe laser beam along with a reference laser beam which were derived from an external cavity diode laser 
operating at $780$ nm. A frequency offset of $800$ MHz between probe and reference beams was introduced using acousto-optic modulators. The coupling laser beam operating in the range of 
$478$ to $482$ counter-propagates the probe beam through the vapor cell. The density of the vapor was varied by heating the cell and the temperature was controlled using a PID controller.
The non-linear phase shift of the probe laser due to two-photon excitation to the Rydberg state in the presence of the coupling laser can be measured by comparing the phase of the reference beam 
using OHDT~\cite{bhow_117}. The details of the OHDT and the theoretical model for relating the dispersion with Rydberg population can be found in reference~\cite{bhow_117}. A typical dispersion spectrum
observed in the experiment is shown in figure 4(b).

\begin{figure}[t]
\includegraphics[scale=0.35]{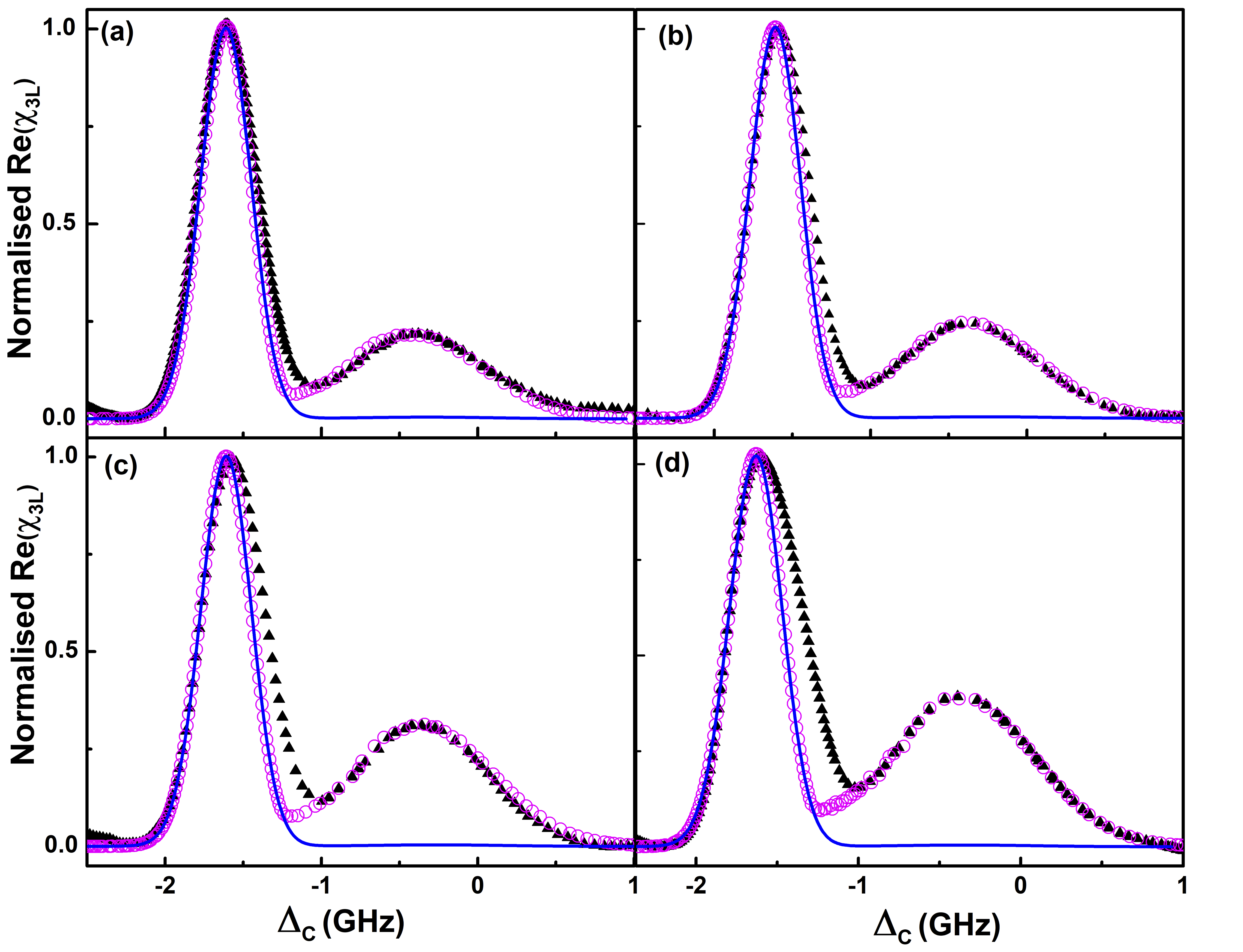}
\caption{\label{fig5} Dispersion spectrum measured from the experiment (black triangle) and calculated from the interacting two-atoms model (\tikzcircle[black]{2pt}) for the Rydberg state with principal
quantum number (a) $n=35$ (b) $n=40$ (c) $n=45$ and (d) $n=53$. For comparison, dispersion calculated from the non-interacting model is depicted as solid lines for all the $n$ states.}
\end{figure} 

The experiment was performed for the Rydberg states $35S_{1/2}$, $40S_{1/2}$, $45S_{1/2}$ and $53S_{1/2}$.  The dispersion of the probe beam was measured using OHDT by varying the density of the rubidium vapor with laser parameters $\Omega_{P}=400$ MHz, $\Omega_{C}=4$ MHz and $\Delta_{P}=1.25$ GHz. All the laser parameters including the gain of the system were kept fixed throughout the experiment for all  Rydberg states. As predicted in the theoretical model, two different peaks were observed for the dispersion spectrum of the probe beam when the coupling laser is scanned over few GHz. One of them is the usual two photon resonant peak and the other one is the anti-blockade peak. Since the Rydberg-Rydberg interaction is repulsive, the anti-blockade peak is expected to be observed on the blue detuned side of the dispersion spectrum. For lower principal quantum number states the Rydberg interactions is weak and is significant only at very high atomic density. However with increase in the principal quantum number of the Rydberg state, the interaction is significant and hence the anti-blockade peak is observed at lower densities as well. The width of the two photon resonant peak is nearly $\Delta kv_p$, while for the anti-blockade peak it is about $k_Cv_p$. The width of the anti-blockade peak is observed to be larger which seems to be in good agreement with the theoretical model as shown in fig. 5. From the expression of $N_b(v_1)$, when the laser is red detuned the contribution of the off resonant atom to the Rydberg population is less as compared to a blue detuned case. Referring to the two photon resonance peak, the anti-blockade effect is significantly larger on the blue detuned side compared to the red detuned side. The two photon resonant peak contains both blockade and anti-blockade effect making it difficult to model. However to have a qualitative understanding, this peak is compared to a non-interacting model. As shown in Fig. 5, with increase in principal quantum number the experimental data deviates from the non interacting model. The deviation on the blue detuned side of the spectrum is an indication of the dominating anti-blockade effect.

\begin{figure}[t]
\includegraphics[scale=0.3]{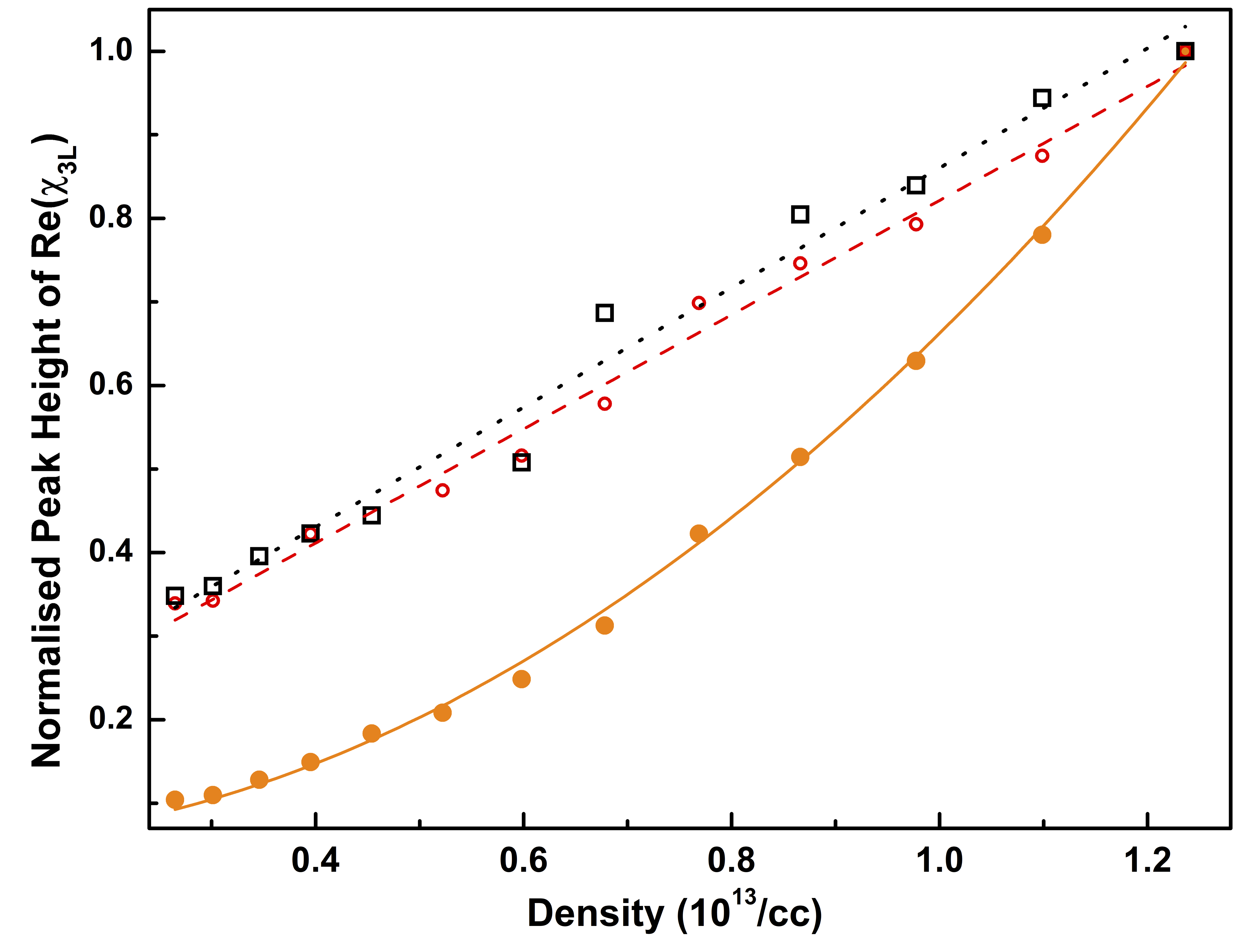}
\caption{\label{fig6} Anti-blockade peak height (\tikzcircle{2pt}) and the usual two-photon resonant peak height for $5$s$_{1/2}$ F$=3\rightarrow n$s$_{1/2}$ (\tikzcircle[red]{2pt}) and 
$5$s$_{1/2}$ F$=2\rightarrow n$s$_{1/2}$ transition($\diamond$) for $n=35$ Rydberg state as a function of the density of the atomic vapor. The peak heights are normalized to the dispersion peak height corresponding to the highest density. Dotted lines are the straight line fitting of usual two-photon 
resonant peaks showing the linear dependence of density whereas the solid line is fitting of the anti-blockade peak height showing quadratic dependance of density.}
\end{figure}
         
For a fixed atomic density, the height of the anti-blockade peak increases with the principal quantum number of the Rydberg state. The $C_6$ scaling with the principal quantum number could not be determined from the experiment. When the number of atoms in the $\ket{g_2}$ state in the interaction sphere is more than one, the blockade effect will contribute along with other cascaded processes involving more atoms. For a Rydberg state with n=35, the interaction is small and the number of atom in the $\ket{g_{2}}$ state in the interaction sphere $N_b\simeq 1$ at a density of $3.0\times 10^{13}/cc$. Thus, the experimental data is expected to match well with the presented model. We measured the dispersion peak height for the anti-blockade peak and the two photon resonance peaks corresponding to the transition $^{85}Rb$ $ F=2\rightarrow 35S_{1/2}$ and $^{87}Rb$ $ F=2\rightarrow 35S_{1/2}$ by varying the density of the vapor cell which is shown in fig. 6. For $35S_{1/2}$, the anti-blockade peak height increases quadratically with increase in density as predicted in the model. For the other two peaks, the variation is observed to be linear. The peak corresponding to the transition $^{85}Rb$ $ F=2\rightarrow 35S_{1/2}$ is expected to be non-interacting as the applied laser is highly detuned from the atomic resonance. So a linear dependence of density is expected. For the two photon resonance peak corresponding to $^{87}Rb$ $ F=2\rightarrow 35S_{1/2}$ the blockade and anti-blockade effect are present which may be compensating each other such that the variation with density is roughly linear. The dotted and dashed lines are the linear fittings and the solid line is the quadratic fit of the peak height data as shown in fig. 6.  

\textbf{Conclusion.}
We have observed the interaction induced enhancement in Rydberg excitation in thermal rubidium vapor. A two atom interacting model is formulated using the dressed state picture of the three level atomic system to explain the anti-blockade peak. The population of the Rydberg state is observed to be enhancing quardratically with the density of the vapor for Rydberg state with n=35, as predicted in the theoretical model. The experiment performed here is limited by the noise in the density measurement. The density dependence of the anti-blockade peak can be studied with a better measurement of density and larger number of data. The deviation from the quadratic behavior can be measured to study the effect of blockade and other cascaded processes on the anti-blockade peak having more than one atom in the $\ket{g_2}$ state in the interaction sphere.

\textbf{Data Availability Statement.}
The data that support the findings of this study are available
from the corresponding author upon request.

\textbf{Acknowledgment.}
We would like to thank Dr. V. Ravi Chandra and Dr. Anamitra Mukherjee for their valuable suggestion for the theoretical modeling. We would like to thank Sushree S. Sahoo, Snigdha S. Pati and Tanim Firdoshi for assisting in performing the experiment. This experiment is financially supported by Department of Atomic Energy, Government of India. 
 
\textbf{Author contributions.}
AKM conceived the idea of anti-blockade in thermal vapor. AKM and DK contributed to the theoretical modeling. DK, AB and AKM contributed in performing the experiment. DK analyzed the experimental data. All authors have contributed to the manuscript preparation.


\begin{thebibliography}{99}
\textbf{{References.}}
\bibitem{saff10}M. Saffman, T. G. Walker, and K. M\o{}lmer Rev. Mod. Phys. 82, 2313 (2010)
\bibitem{dudi_112} Y. O. Dudin, L. Li, F. Bariani and A. Kuzmich, Nature Physics 8, 790–794 (2012)
\bibitem{dudi_212} Y. O. Dudin, and A. Kuzmich, Science 336, 887 (2012)
\bibitem{tong04} D. Tong, S. M. Farooqi, J. Stanojevic, S. Krishnan, Y. P. Zhang, R. C\^ot\'e, E. E. Eyler, and P. L. Gould, Phys. Rev. Lett. 93, 063001 (2004)
\bibitem{sing04} K. M. Singer, M. Reetz-Lamour, T. Amthor, L. G. Marcassa, and M. Weidem\"uller, Phys. Rev. Lett. 93, 163001(2004)
\bibitem{lieb05} T. Cubel Liebisch, A. Reinhard, P. R. Berman, and G. Raithel, Phys. Rev. Lett. 95, 253002 (2005)
\bibitem{vogt06} T. Vogt, M. Viteau, J. Zhao, A. Chotia, D. Comparat, and P. Pillet, Phys. Rev. Lett. 97, 083003 (2006)
\bibitem{heid07} R. Heidemann, U. Raitzsch, V. Bendkowsky, B. Butscher, R. L\"ow, L. Santos, and T. Pfau, Phys. Rev. Lett. 99, 163601 (2007)
\bibitem{rait08} U. Raitzsch, V. Bendkowsky, R. Heidemann, B. Butscher, R. L\"ow and T. Pfau, Phys. Rev. Lett. 100, 013002 (2008)
\bibitem{urba09} E. Urban, T. A. Jojnson, T. Henage, L. Isenhower, D. D. Yavuz, T. G. Walker, and M. Saffman, Nature Phys. 5, 110 (2009)
\bibitem{gaet09} A. Ga\"etan, Y. Miroshnychenko, T. Wilk, A. Chotia, M. Viteau, D. Comparat, P. Pillet, A. Browaeys, and P. Grangier, Nature Phys. 5, 115 (2009)
\bibitem{weim08} H. Weimer, R. L\"ow, T. Pfau, and H. P. B\"uchler, Phys. Rev. Lett. 101, 250601 (2008)
\bibitem{pohl10} T. Pohl, E. Demler, and M. D. Lukin, Phys. Rev. Lett. 104, 043002 (2010)
\bibitem{pupi10} G. Pupillo, A. Micheli, M. Boninsegni, I. Lesanovsky, and P. Zoller, Phys. Rev. Lett. 104, 223002 (2010)
\bibitem{henk10} N. Henkel, R. Nath, and T. Pohl, Phys. Rev. Lett. 104, 195302 (2010)
\bibitem{busc17} Hannes Busche, Paul Huillery, Simon W. Ball, Teodora Ilieva, Matthew P. A. Jones and Charles S. Adams, Nature Physics 13, 655–658 (2017)
\bibitem{peyr12} Thibault Peyronel, Ofer Firstenberg, Qi-Yu Liang, Sebastian Hofferberth, Alexey V. Gorshkov, Thomas Pohl,Mikhail D. Lukin and Vladan Vuleti\'c, Nature 488, 57–60 (2012)
\bibitem{fist13} Ofer Firstenberg, Thibault Peyronel, Qi-Yu Liang, Alexey V. Gorshkov, Mikhail D. Lukin1 and Vladan Vuleti\'c, Nature 502, 71–75 (2013) 
\bibitem{jaks00} D. Jaksch, J. I. Cirac, P. Zoller, S. L. Rolston, R. C\^ot\'e and M. D. Lukin, Phys. Rev. Lett 85, 2208 (2000)
\bibitem{isen10} L. Isenhower, E. Urban, X. L. Zhang, A. T. Gill, T. Henage, T. A. Johnson, T. G. Walker and M. Saffman, Phys. Rev. Lett. 104, 010503 (2010)
\bibitem{wilk10} T. Wilk, A. Ga\"etan, C. Evellin, J. Wolters, Y. Miroshnychenko, P. Grangier, and A. Browaeys, Phys. Rev. Lett. 104, 010502 (2010)
\bibitem{luki01} M. D. Lukin, M. Fleischhauer, R. Cote, L. M. Duan, D. Jaksch, J. I. Cirac, and P. Zoller, Phys. Rev. Lett. 87, 037901 (2001)
\bibitem{ates07} C. Ates, T. Pohl, T. Pattard and J. M. Rost, Phys. Rev. Lett. 98, 023002 (2007)
\bibitem{amth10} T. Amthor, C. Giese, C. S. Hofmann and M. Weidmuller, Phys. Rev. Lett. 104, 013001 (2010)
\bibitem{pohl09} T. Pohl and P. R. Berman, Phys. Rev. Lett. 102, 013004 (2009)
\bibitem{qian09} J. Qian, Y. Qian, M. Ke, X. L. Feng, C. H. Oh, and Y. Z. Wang, Phys. Rev. A 80, 053413 (2009)
\bibitem{sark15} L. S\'ark\'any, J. Fort\'agh, and D. Petrosyan, Phys. Rev. A 92, 030303(R) (2015)
\bibitem{su16} S. L. Su, Erjun Liang, S. Zhang, J. J. Wen, L. L. Sun, Z. Jin, and A. D. Zhu, Phys. Rev. A 93, 012306 (2016)
\bibitem{su17} Shi-Lei Su, Ya Gao, Erjun Liang and Shou Zhang, Phys. Rev. A 95, 022319 (2017)
\bibitem{balu13} T. Baluktsian, B. Huber, R. L̈\"ow, and T. Pfau, Phys. Rev. Lett. 110, 123001 (2013)
\bibitem{klei17}K. S. Kleinbach, F. Meinert, F. Engel, W. J. Kwon, R. L\"ow, T. Pfau, and G. Raithel, Phys. Rev. Lett. 118, 223001 (2017)
\bibitem{melo16}Natalia R. de Melo, Christopher G. Wade, Nikola \ifmmode \check{S}\else \v{S}\fi{}ibali\ifmmode \acute{c}\else \'{c}\fi{}, Jorge M. Kondo, Charles S. Adams and Kevin J. Weatherill, Phys. Rev. A 93, 063863 (2016)
\bibitem{kubl10} H. K̈\"ubler, J. P. Shaffer, T. Baluktsian, R. L\"ow, ̈T. Pfau, Nature Phot. 4, 112 (2010)  
\bibitem{moha07} A. K. Mohapatra, T. R. Jackson, and C. S. Adams, Phys. Rev. Lett. 98, 113003 (2007)
\bibitem{koll12} A. K̈\"olle, G. Epple, H. K ̈ubler, R. L\"ow and T. Pfau, Phys. Rev. A 85, 063821 (2012)
\bibitem{bhow16} Arup Bhowmick, Sushree S. Sahoo, and Ashok K. Mohapatra Phys. Rev. A 94, 023839 {2016}     
\bibitem{bhow_217} Arup Bhowmick, Dushmanta Kara, Ashok K. Mohapatra, Manuscript Under Preparation 
\bibitem{lets17}F. Letscher, O. Thomas, T. Niederpr\"um, H. Ott, and M. Fleischhauer, Phys. Rev. A 95, 023410, (2017)
\bibitem{bhow_117} Arup Bhowmick, Dushmanta Kara, Ashok K. Mohapatra, arXiv:1709.07750v1
\bibitem{begu13} L. B\'eguin, A. Vernier, R. Chicireanu, T. Lahaye, and A. Browaeys, PRL 110, 263201 (2013)
\end{thebibliography}

\end{document}